\documentclass[prb,twocolumn,showpacs,amsmath,amssymb,superscriptaddress]{revtex4-1}
\usepackage{graphicx}
\usepackage{dcolumn}
\usepackage{bm}
\usepackage{color}
\usepackage{hyperref}
\hypersetup{backref=true,
pdfnewwindow=true, colorlinks=true,
linkcolor=blue, anchorcolor=blue,
citecolor=blue, filecolor=blue,
menucolor=blue, urlcolor=blue}
\def\bal {\begin{align}}
\def\eal {\end{align}}
\def\be {\begin{equation}}
\def\ee {\end{equation}}
\def\bea {\begin{eqnarray}}
\def\eea {\end{eqnarray}}

\def\ob {\Big{[}} 
\def\cb {\Big{]}}
\def\ep {\varphi}
\def\dk {\delta}

\def\beq{\begin{equation}}\title{Some title}
\def\eeq{\end{equation}}
\begin{document}
\title{Approximations for many-body Green's functions: insights from the fundamental equations}
\author{Giovanna Lani}
\address{
Laboratoire des Solides Irradi\'es, Ecole Polytechnique, CNRS,CEA-DSM, F-91128 Palaiseau, France}
\address{
European Theoretical Spectroscopy Facility (ETSF)
}
\author{Pina Romaniello}
\address{
Laboratoire des Solides Irradi\'es, Ecole Polytechnique, CNRS,CEA-DSM, F-91128 Palaiseau, France}
\address{
European Theoretical Spectroscopy Facility (ETSF)
}
\address{
Laboratoire de Physique Th\'{e}orique-IRSAMC, CNRS,
Universit\'{e} Paul Sabatier, F-31062 Toulouse, France
}
\author{Lucia Reining}
\address{
Laboratoire des Solides Irradi\'es, Ecole Polytechnique, CNRS,CEA-DSM, F-91128 Palaiseau, France}
\address{
European Theoretical Spectroscopy Facility (ETSF)
}

\date{\today}

%

\begin{abstract}
Several widely used methods for the calculation of band structures and photo emission spectra, such as the GW approximation, rely on Many-Body Perturbation Theory. They can be obtained by iterating a set of functional differential equations relating the one-particle Green's function to its functional derivative with respect to an external perturbing potential.
In the present work we apply a linear response expansion in order to obtain insights in various approximations for Green's functions calculations. The expansion leads to an effective screening, while keeping the effects of the interaction to all orders. In order to study various aspects of the resulting equations we discretize them, and retain only one point in space, spin, and time for all variables. Within this one-point model we obtain an explicit solution for the Green's function, which allows us to explore the structure of the general family of solutions, and to determine the specific solution that corresponds to the physical one. Moreover we analyze the performances of established approaches like $GW$ over the whole range of interaction strength, and we explore alternative approximations. Finally we link certain approximations for the exact solution to the corresponding manipulations for the differential equation which produce them. This link is crucial in view of a generalization of our findings to the real (multidimensional functional) case where only the differential equation is known.
\end{abstract}
\maketitle
%
%

\section{Introduction}
\label{sec:intro}

The one-particle Green's function (GF) \cite{fetterwal,abrikosov,Hedin_Lundqvist} is a powerful quantity since it contains a wealth of information about  a physical system, such as the expectation value of any single-particle operator over the ground state, the ground-state total energy, and the spectral function. In order to access this quantity one can start from its equation of motion: \cite{kadanoffbaym,Strinati,Csanak}
\begin{eqnarray}
\displaystyle
 \label{eq:eom}
&\Big{[} & i \frac{\partial}{\partial t_1} - h(r_1) \Big{]} G(1,2)  \nonumber \\
&+& i\int d3 \: v(1^+,3) G_2(1,3;2,3^{+}) = \delta(1,2), 
\label{Eqn:EOM}
\end{eqnarray}
where $h(r_1)$ is the one-electron part of the many-body Hamiltonian, $G_2(1,3;2,3^{+})$ is the two-body Green's function, and $v(1^{+},3)$ is the Coulomb potential. The space, spin and time variables are all combined in $(1)=\displaystyle  (r_1,\sigma_1,t_1) $,  and $\displaystyle (1^+)=(r_1,\sigma_1,t^+_1)$ with $t_1^+=t_1+\delta$ ($\delta\rightarrow 0^+$). \\
Eq.~(\ref{Eqn:EOM}) can be manipulated in order to get a more practical expression, by introducing the non-interacting Green's function $G_0$ with%
\begin{eqnarray}
\label{eq:g0}
\displaystyle 
\ob  i\frac{\partial }{\partial t_1} - h(r_1)\cb G_0(1,2) = \delta(1,2),
\end{eqnarray}
which reinserted in Eq.~(\ref{eq:eom}), gives
\begin{eqnarray}
 \label{eq:dyson}
G(1,2) &=& G_0(1,2) \nonumber \\
&-& i\int d3 d4 G_0(1,3) v(3^{+},4)G_2(3,4;2,4^{+}).
\end{eqnarray}
In~(\ref{eq:dyson}) $G_0$ determines the appropriate initial condition in time; note that the solutions of (\ref{Eqn:EOM}) and (\ref{eq:g0}) are not unique. Moreover, in order to calculate $G$,  the knowledge of $G_2$ is required (which in turns requires the knowledge of $G_3$, and so on) \cite{kadanoffbaym,Csanak}. In order to obtain a closed expression one can generalize $G(1,2)$ to $G(1,2;[\varphi])$, where an external fictitious time-dependent potential $\varphi$ is applied to the system. This allows one to express $G_2$ as ~\cite{martsch59}
\begin{eqnarray}
 \label{eq:schw}
 \displaystyle
 G_2(3,4;2,4^{+};[\varphi]) &=&  G(3,2;[\varphi])G(4,4^{+};[\varphi]) \nonumber \\
&-& \frac{\delta G(3,2;[\varphi])}{\delta \varphi(4)}.
 \label{Eqn:Schwinger}
 \end{eqnarray}  
 Note that in~(\ref{Eqn:Schwinger}) all Green's functions are
 generalized to non-equilibrium since they depend on the perturbing potential.
 The equilibrium $G$ and $G_2$ in (\ref{eq:dyson}) are then obtained by taking $\varphi=0$. Inserting (\ref{Eqn:Schwinger}) into (\ref{eq:dyson}) yields a set of functional differential equations \cite{kadanoffbaym} for the unknown $G$
\begin{eqnarray}
 \label{eq:diffnd} 
&&G (1,2;[\varphi]) = G_0 (1,2) \nonumber \\
+&& \int d3 \:  G_0(1,3)V_H(3;[\varphi])G(3,2;[\varphi]) \nonumber \\
+&& \: \int d3 \: G_0(1,3)\varphi(3) G(3,2;[\varphi]) \nonumber \\
+&&  \: i \int d4 d3 \:  G_0(1,3) v(3^{+},4)\frac{\delta G(3,2;[\varphi])}{\delta \varphi(4)},  
\end{eqnarray}
where $\displaystyle V_H(3)= -i\int d4 \: v(3,4)G(4,4^{+};[\varphi])$ is the Hartree potential. Since the Hartree potential contains the Green's function, this term makes the equations nonlinear. We are interested in the solution of Eq.~(\ref{eq:diffnd}), for $\varphi=0$.
Its calculation would hence require the solution of a set of coupled, non-linear, first-order differential equations, which is clearly a non trivial task. Moreover one would need a new initial condition to completely define the desired solution of this differential equation, since the derivative $\displaystyle \frac{\delta G}{\delta \varphi} $ has been introduced. Therefore usually another route is taken: one includes the functional derivative in (\ref{eq:diffnd}) in the definition of a self-energy \cite{kadanoffbaym}
\begin{eqnarray}
\label{eq:defsig}
 \displaystyle  
\Sigma(1,3) &=& i \int d4 d2 \: v(1^{+},4) \frac{\delta G(1,2;[\varphi])}{\delta \varphi(4)}\Big{|}_{\varphi = 0} \nonumber \\
&\times& G^{-1}(2,3),
\end{eqnarray}
which, inserted into Eq.~(\ref{eq:diffnd}) for $\varphi = 0 $, gives:
\begin{eqnarray}
 \label{eq:dysonsig}
G(1,2) &=&  G_0(1,2) + \int d3 \:  G_0(1,3)V_H(3)G(3,2)   \nonumber \\
&+& \int d4 d3 \: G_0(1,3) \Sigma(3,4) G(4,2). 
\end{eqnarray}
This is the Dyson equation for $G$, where $\Sigma$ contains all the many-body effects (beyond the Hartree contribution) present in the system. Of course $\frac{\delta G}{\delta\varphi}$ and therefore $\Sigma$ are still not known and, in practice, $\Sigma$ has to be approximated. A good starting point is obtained by reformulating the problem in terms of a coupled set of equations  containing the one-particle Green's function, the \textit{polarizability} $P$, the \textit{self-energy} $\Sigma$, the \textit{screened Coulomb interaction} $ W$, and the \textit{vertex} $\Gamma$. 
These equations are most often solved within the so called $GW$ approximation ($GW$A) \cite{hedin65}, where the vertex $\Gamma$ is set to unity, resulting in $\displaystyle \Sigma \approx iGW$. Over the last two decades the $GW$ method has become the tool of choice for calculations of quasi-particles (QP) band structures~\cite{Aulbur,Gunnarsson} of many materials and direct and inverse photo emission spectra (see e.g. Ref.~\cite{gatti-VO2,gatti-V2O3, Kotani07, Kotani04}) improving substantially over the results provided by static mean-field electronic structure methods. \\
However the $GWA$ suffers from some fundamental shortcomings (see e.g. Refs.~\cite{von-Barth,Godby,Aryasetiawan-Tmatrix,pina09}) and with $\Sigma$ being of first order in $W$, is not expected to describe strong correlation. Higher orders in $W$ could be added by iterating the equations, but this is technically difficult, and there is no guarantee that results will quickly improve. It is therefore necessary to find guidelines.\\
In the present work we go back to Eq.~(\ref{eq:diffnd}). Our aim is first, to obtain new insights about standard approximations by relating them more directly to the original equations. Second, we want to use   Eq.~(\ref{eq:diffnd}) to explore alternative approximations. Finally, it might be interesting to concentrate directly on the set of coupled, non linear, first order functional differential equations for $G$, Eq~(\ref{eq:diffnd}), although it has been acknowledged that no "practical technique for solving such functional differential equation exactly" \cite{kadanoffbaym}  is available. However, one may still hope that with new algorithms and the increase in computer power, numerical solutions might become accessible. The present work is hence also meant to explore strategies for, and possible problems of, such a route. 
In the following we resort to two approximations. First we linearize the set of equations by expanding $V_H$ in terms of $\varphi$. Second, we discretize Eq.~(\ref{eq:diffnd}) and consider in a first instance only one point for each space, spin, and time variable: we will call this latter approximation the \textit{"1-point model"}, as opposed to the full functional problem. 
The strategy underlying this procedure is the following: for the 1-point model, we can derive the exact explicit solution of the now algebraic differential equation, and solve the initial value problem. One can hence explore approximations to the full solution, which yields valuable insights in the performance of current approaches and suggestions for alternative ones. By determining which manipulations of the differential equation (DE) produce such approximate solutions, one obtains suggestions for analogous  manipulations on the differential equation for the full functional problem, which opens the way to translate our model findings to realistic calculations.\\

\noindent The manuscript is structured as follows. In Section~\ref{sec:eq} we present the linearized differential equation which can be solved exactly within the 1-point framework. We discuss in particular the initial value problem, and how it can be overcome. In Section~\ref{sec:benchmark} we examine, in the 1-point framework, various common approximations to the solution of the DE: the iteration of Eq.~(\ref{eq:diffnd}) and approximations based on a Dyson equation, in particular different $GW$ flavours.
In Section~\ref{sec:newmethod} we explore other routes to manipulate the initial DE and obtain approximate solutions. 
We finally give our conclusions and perspectives on future work in Section~\ref{sec:concl}.
\section{The screened equation in a 1-point framework} 
\label{sec:eq}
\begin{figure}
 \includegraphics[width=2.5in,angle=360]{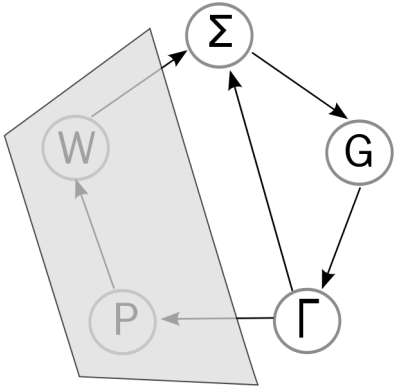}
\caption{Hedin's pentagon when $W$ is kept fixed: one iterates only three equations, namely the ones for $G, \Sigma, \Gamma$ rather than the full set. Note that keeping $W$ fixed also implies fixing the polarizability $P$.}
\label{fig:8}
\end{figure}

Our first goal is to simplify the equations such that the main physics is retained, but manipulations become more straightforward. To this end we linearize the differential equation with an expansion of the Hartree potential to first order in the external potential $\varphi$, 
\bea
 \displaystyle
&& V_H(3;[\varphi]) \approx -i \int d4 v(3^+,4) G(4,4^{+};[\varphi]) \Big{|}_{\varphi = 0} \nonumber \\
&& - i\int d4 \,d5 \,v(3^+,4) \frac{\delta G(4,4^{+};[\varphi])}{\delta \varphi(5)}\Big{|}_{\varphi = 0} \varphi(5)  + o(\varphi^2).
\eea
Eq.~(\ref{eq:diffnd}) hence becomes
\bea
 \label{eq:nohart}
G(1,2;[\bar{\varphi}]) &=& G_H^0(1,2) +  \int d3  G_H^0(1,3) \bar{\varphi}(3) G(3,2;[\bar{\varphi}]) \nonumber \\
&+& i \! \int d3 d5 G_H^0(1,3) W(3^+,5) \frac{\delta G(3,2;[\bar{\varphi}]) }{\delta\bar{\varphi}(5)},
\label{Eqn:linerized_ODE}
\eea
where $G_H^0$ is a Hartree Green's function containing the Hartree potential at vanishing $\varphi$,  $\displaystyle \bar{\varphi}=\epsilon^{-1}\varphi$ is the renormalized external potential, and $\displaystyle W=\epsilon^{-1}v$ is the screened Coulomb potential  with $\epsilon$ the dielectric function at $\varphi=0$ \footnote{For simplicity we use the same symbol for $G[\bar{\varphi}]$ and $G[\varphi]$; of course it is understood that the corresponding functional is taken}.\\
Concerning Eq.~(\ref{Eqn:linerized_ODE}) three important remarks should be made: first, through the linearization the screened interaction $W$ becomes the central quantity of the equation. This is justified by the physics of extended systems, where screening and plasmons are key concepts. \\
Second, in principle $W$ is the exact screened interaction, which of course is not known.
One can however adopt two strategies: either $W$ is considered to be an \textit{externally} given quantity, obtained within a good approximation, e.g. from a time dependent density functional theory (TDDFT) calculation~\cite{runge-dft}; or one could also recalculate $W$ from $G[\bar{\ep}]$ (see in the next section). In this work we will adopt the first strategy, which is illustrated in Fig.~\ref{fig:8}. 
Such a philosophy is rigorously justified. In particular, in the framework of the theory of functionals it is possible to pass from the Luttinger-Ward functional (given as functional of G, though indeed one should add the bare Coulomb interaction $v$ as argument) to the so-called $\Psi$-functional, where $v$ is replaced by $W$ ~\cite{ulf-05}. 
This explains for example why self-consistency in G only (and not in W) is sufficient to have a conserving GW approximation. 
Moreover, in practice this is the most current way of proceeding, corresponding e.g. in a GW calculation to the "best G, best W" approach: while the non-interacting G is taken e.g. to be the Kohn-Sham Green's function, $W$ is calculated as accurately as possible, e.g. in the adiabatic local density approximation to TDDFT. 
Third by approximating the functional derivative $\displaystyle \frac{\delta G}{\delta \bar{\varphi}}=\displaystyle -G \frac{\delta G^{-1}}{\delta\bar{ \varphi}} G\approx GG$ (which supposes the self-energy to be independent of ${\bar{\varphi}}$) one obtains the Dyson equation for the one-particle Green's function in the $GW$ approximation to the self-energy. The proof is given in App.~\ref{sec:appahar}. This result shows that, even though the linearization procedure is an approximation, Eq.~(\ref{Eqn:linerized_ODE}) is still a promising starting point to analyze the different flavours of the $GW$ approximations and to go beyond.\\
After linearizing, the next step consists in \textit{discretizing}  Eq.~(\ref{Eqn:linerized_ODE}) and then in considering only one value for the space, spin, and time variables respectively (or equivalently concerning space and spin, in considering all Green's functions to be diagonal in a given basis): this is the 1-point model employed throughout the whole manuscript. The 1-point framework has already been used by other authors: in Refs~\cite{molinari,molinari06} Hedin's equations are combined  in one single algebraic differential equation which is solved as a series expansion. This allows the authors to enumerate the diagrams for a certain order of expansion. Several expansion parameters are examined, such as $vG_H^2$, with $v$ the bare Coulomb potential and $G_H$ the Hartree Green's function, $vG^2$, with $G$ the exact  Green's function, $WG^2$, with $W$ the screened Coulomb potential, etc., which shows how at various orders of expansion the number of diagrams decreases by increasing the degree of renormalization.  This is also the spirit behind the linearized equation (\ref{Eqn:linerized_ODE}), in which the natural expansion parameter would be $WG_H^2$, where $W$ is treated as an externally given interaction. The advantage of using the 1-point framework is that the equations become algebraic and thus the enumeration of diagrams is facilitated.
In Ref.~\cite{hubner} a similar strategy as in Refs ~\cite{molinari,molinari06} is used to enumerate diagrams, focusing in particular on the asymptotic behavior of the counting numbers. Moreover Hedin's equations are transformed into a single first oder differential equation for the GF as a function of an interaction parameter, and an implicit solution is obtained. In order to fix the particular solution of this differential equation the initial condition $G_{(v=0)}=G_0$ is used.
\noindent Instead here we concentrate on (\ref{eq:diffnd}), or better its linearized form (\ref{Eqn:linerized_ODE}), which is \textit{another} differential equation for $G$, as a functional of an external potential. This choice allows us to \textbf{(i)} emphasize the essential physics contained in the screened Coulomb interaction $W$, \textbf{(ii)} discuss various aspects of the many-body problem in a clear and simple way, \textbf{(iii)} obtain an exact solution of the approximate equation that can be used as a benchmark. Moreover we believe that the 1-point version of Eq.~(\ref{Eqn:linerized_ODE}) can be a natural starting point for a generalization to the full functional problem.
While the equations are easier to manipulate, physical information is of course lost in the 1-point framework. In particular no poles (addition/removal energies) of the GF appear. However, the various aspects that will be explored in the following are intrinsically related to the structure of the equations, and hence exportable also to the full functional problem, in the same spirit as in Refs~\cite{molinari,molinari06,hubner}.

%
%
\subsection{The 1-point differential equation}
In the 1-point model Eq.~(\ref{eq:diffnd}) reduces to an algebraic, non-linear, first order differential equation
\be
\displaystyle
\label{Eqn:diff0dfull}
y(z) =  y_0  +v y_0 y^{2}(z) + y_0 zy(z) -v y_0 \frac{d \: y(z)}{d z} 
\ee 
where 
$\displaystyle\ \varphi \overset{\underset{\mathrm{}}{}}{\rightarrow}  z$, $G(1,2;[\varphi]) \overset{\underset{\mathrm{}}{}}{\rightarrow} y(z)$, and
$G_0(1,2) \overset{\underset{\mathrm{}}{}}{\rightarrow}  y_0$. Moreover $i v(3^{+},4) \overset{\underset{\mathrm{}}{}}{\rightarrow} -v$: this change of prefactor compensates for the time- or frequency integrations that have been dropped in the 1-point model and corresponds to a standard procedure in this context~\cite{molinari,hubner}. 
We can now linearize Eq.~(\ref{Eqn:diff0dfull}) in the same way as we did starting with  Eq.~(\ref{eq:diffnd}) and obtaining Eq.~(\ref{Eqn:linerized_ODE}). This yields
\be
 \label{Eqn:diff0dnotr} 
y_u(x) =  y^{0}_H  + y^0_H x y_u(x) -u  y^0_H \frac{d \: y_u(x)}{d x}.  
\ee
Hence with respect to Eq.~(\ref{Eqn:linerized_ODE}), $\displaystyle\ \bar{ \varphi} \overset{\underset{\mathrm{}}{}}{\rightarrow}  x$, $G_H^0(1,2) \overset{\underset{\mathrm{}}{}}{\rightarrow}  y^H_0$, and  $i W(3^{+},5) \overset{\underset{\mathrm{}}{}}{\rightarrow} -u$; the subscript $u$ in $y_u$ highlights its $u$ dependence. In the following, for simplicity of notation, we denote $y^{0}_H$ by $y_0$ unless stated differently. 
In Appendix~\ref{sec:appab} we sketch the main steps to solve Eq.~(\ref{Eqn:diff0dnotr}), based on the general ansatz $y_u(x)=A(x)\cdot \mathcal{I}(x)$. With the choice
\begin{eqnarray}
 \displaystyle
A(x) =  e^{\left[ \frac{x^{2}}{2u} - \frac{x}{uy_0} \right] }
\end{eqnarray}
one obtains the equation 
\begin{eqnarray}
 \displaystyle
\label{Eqn:findI}
 \frac{d \mathcal{I}(x)}{d x} = \frac{1}{u}e^{-\left[ \frac{x^{2}}{2u} - \frac{x}{uy_0}\right] }
\end{eqnarray}
and the general solution $\displaystyle y_u(x)$ reads
\bea
 \label{eq:gensolode_2} 
y_u(x) &=& \sqrt{ \frac{\pi}{2u} } e^{ \left[ \frac{x^{2}}{2u}  - \frac{x}{uy_0} + \frac{1}{ 2uy_0^{2}} \right] } \nonumber \\  
& \times& \left\{ erf \left[ (x -\frac{1}{y_0})\sqrt{\frac{1}{2u}} \right] - C(y_0,u)  \right\},
\label{Eqn:linear_1point_ODE}
\eea
where $C(y_0,u)$ is to be set by an initial condition. 
In the limit $x \rightarrow 0$, which is the equilibrium solution we are looking for, Eq.\ (\ref{Eqn:linear_1point_ODE}) becomes
\bea
\label{eq:gensolode}
\displaystyle
y_u &=& - \sqrt{\frac{\pi}{2u}} \: e^{\frac{1}{2u y_0^{2} }} \nonumber \\
&\times & \left \{ erf \ob \sqrt{\frac{1}{2uy_0^{2}}} \cb + C (y_0,u) \right \}.
\eea
Note that a similar ansatz can also be used  for the full functional problem, namely $\displaystyle G(1,2) = \int d3 \, A(1,3) \cdot I(3,2)$, in order to get a set of differential equations that are less complicated to manipulate than the original one.
\subsection{The initial value problem\label{sec:iniva}}
In general in order to set $C(y_0,u)$, $y_u(x)$ has to be known for a given potential $x_\beta$ (i.e. $\displaystyle y_u( x_{\beta }) = y_u^{\beta} $). However it is far from obvious to formulate such a condition in the realistic full functional case; this would indeed require the knowledge of the full interacting $G$ for some given potential $\varphi$. Therefore the question is whether one can reformulate the condition in a simpler way in order to set $C$. \\
To answer this question we expand the exact solution for small values of $u$, obtaining: 
\begin{eqnarray}
\label{eq:odexp}
 y_u &\approx& - \sqrt{\frac{\pi}{2u}} e^{ \frac{1}{2uy_0^{2} } } \Big{(}1+C(u,y_0)\Big{)} \nonumber \\
     & + & \Big{\{} y_0 - uy_0^{3}  + 3u^{2}y_0^{5} - 15u^{3} y_0^{7} + o(u^{4}) \Big{\}} .
\end{eqnarray}
When $u \rightarrow 0$ the one-body Green's function $G$ has to reduce to the non interacting $G_0$, in our framework this translates into: $\displaystyle \label{eq:xulimit} y_u\Big{|}_{u \rightarrow 0} \equiv y_0 $. Imposing this condition on Eq.~(\ref{eq:odexp}) gives
\begin{eqnarray}
 \label{eq:condit}
 \sqrt{\frac{\pi}{2u}} e^{\ob \frac{1}{2y_0^{2}u} \cb} \Big{(} 1+C(u,y_0) \Big{)} = 0, \quad \quad u \rightarrow 0,
\end{eqnarray}
which is satisfied if
\begin{eqnarray}
 \label{eq:fixingc}
C(u,y_0) = -1, \quad \quad u \rightarrow 0.
\end{eqnarray}
This result for $C$ holds also for $u \neq 0$. Indeed it guarantees a non-divergent result for any non-vanishing potential $x$ in (\ref{Eqn:linear_1point_ODE}). Moreover it reproduces the perturbative result which is obtained by iterating Eq.~(\ref{Eqn:diff0dnotr}); for example the sixth iteration yields
\bea
  \label{eq:exp0d}
   y_u^{(6)} = y_0 - u y_0^{3} +  3u^{2} y_0^{5} - 15u^{3} y_0^{7}.
\eea

This is precisely the same series as the one appearing in Eq.~(\ref{eq:odexp}) when $C(u,y_0)$ is set to $-1$.
%
%
%

%
\section{Analysis of common methods to calculate the one-body G}
\label{sec:benchmark}
In the following we will analyze various established approximations for the calculation of the one particle $G$, using the knowledge of the exact solution. %
\subsection{Iteration of the DE}
\label{subsec:valid}

\begin{figure}
\includegraphics[width=3.4in,angle=360]{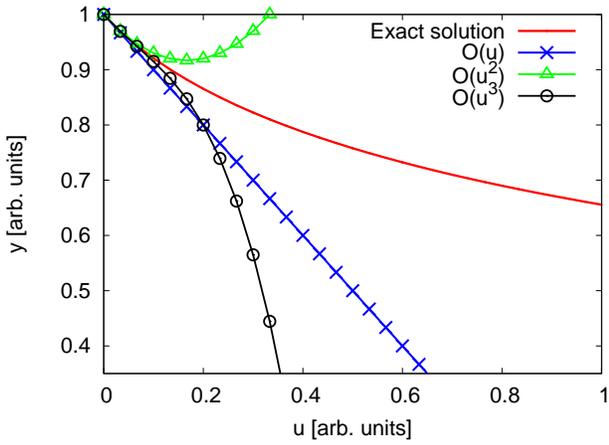}
\caption{Comparison between the exact solution (red plain line, Eq.~(\ref{eq:gensolode})) and the iterative solution for $x=0$ of (Eq.~(\ref{Eqn:iteration})). The blue crosses represent the $1^{st}$ order expansion (Eq.~(\ref{Eqn:3orders_de_1})), while the green triangles and the black circles are respectively the $2^{nd}$ (Eq.~(\ref{Eqn:3orders_de})) and $3^{rd}$ order (Eq.~(\ref{eq:exp0d})). All the three orders are close to the exact solution for small $u$ values, whereas when a given order of the series starts to diverge, the lower orders of the expansion reproduce the exact results better. For each curve $C(u,y_0)= -1$, and we arbitrarily set $y_0 = 1$.}
\label{fig:1}
\end{figure}

Let us first iterate Eq.~(\ref{Eqn:diff0dnotr}) starting from $\displaystyle y_u^{(0)}(x) = y_0$, according to:
 \bea
  y_u^{(n+1)}(x) = y_0 + y_u^{(n)}xy_0 -uy_0 \frac{d y_u^{(n)}(x)}{dx}.
  \label{Eqn:iteration}
 \eea
 For $x=0$ the first two orders in $u$ read 
\bea
\label{Eqn:3orders_de_1}
  y_u^{(2)} &=& y_0 - u y_0^{3},  \\
  y_u^{(4)} &=& y_0 - u y_0^{3} +  3u^{2} y_0^{5}, 
 \label{Eqn:3orders_de}
\eea
and Eq.~(\ref{eq:exp0d}) for the third order. Results as a  function of $u$ are depicted in Fig.~\ref{fig:1} together with the exact solution. 
Two observations can be made: \textbf{i)} very few terms are needed to obtain a good approximation to the exact solution in the small $u$ regime; \textbf{ii)} for a given $u=u_n$, the expansion diverges starting from an order $n$. The larger is $u_n$, the smaller is $n$, which limits the precision that can be obtained.  
As previously mentioned, the iteration coincides with the expansion for small $u$ of the exact solution. Since the small $u$ expansion is  \textit{de facto} the asymptotic expansion of the error function times an exponential (as can be seen in (\ref{eq:odexp}))  the divergent behaviour of the iteration in (\ref{Eqn:iteration}) is not surprising.
Divergences of higher orders have been found in perturbation expansions for realistic systems, e.g. for orders higher than 3 in the  M$\o$eller-Plesset scheme  \cite{leininger00,dunning98}.
%

\subsection{Self-energy based approximations}

In this section the introduction of a self-energy $\Sigma$ will be discussed along with its most common approximations.\\ 
The Dyson-like form for Eq.~(\ref{Eqn:diff0dnotr}), which is the equivalent of Eq. (\ref{eq:dysonsig}), reads:
\begin{eqnarray}
\label{eq:dyson2}
y_u(x)= y_{0} + y_0x y_u(x) + y_0 \Sigma_u \left [ y_u(x) \right ] y_u(x)
\label{Eqn:Dyson_eq}
\end{eqnarray}
where a self-energy kernel has been defined as
\bea
 \displaystyle
 \label{eq:defsig}
 \Sigma_u \left[ y_u(x) \right] &=&  -u \frac{d y_u(x)}{d x} \frac{1}{y_u(x)}.
\eea
With $\displaystyle \frac{dy_u(x)}{dx} = -y^2_u(x) \frac{dy_u^{-1}(x)}{dx}$ and the definition $\displaystyle \Gamma_u \left[ y_u(x) \right] = -\frac{dy_u^{-1}(x)}{dx}$ for the vertex function, the self-energy reads 
\bea
 \label{eq:sigmad}
 \Sigma_u \left[ y_u(x) \right] &=& -u y_u(x) \Gamma_u \left[ y_u(x) \right],
\eea
which is the equivalent of $\displaystyle \Sigma=iGW \Gamma$ \cite{hedin65}. The Bethe-Salpeter equation for the vertex function $\Gamma$ is then derived from~(\ref{Eqn:Dyson_eq}) 
\begin{eqnarray}
 \frac{dy_u^{-1}(x)}{dx} &=& -1-\frac{d \Sigma_u \left[ y_u(x) \right] }{dx} \nonumber \\
&=& -1- \frac{d \Sigma_u \left[   y_u(x) \right] }{d y_u(x)} \frac{d y_u(x)}{dx},
\end{eqnarray}
from which for $x\to 0$
\bea
 \displaystyle
 \label{eq:2hedin}
 \Gamma_u(y_u)  & = & 1 + \frac{d \Sigma_u (y_u)}{d y_u} \Gamma_u(y_u) y_u^{2}
\eea
where $\displaystyle y_u = y_u (x\to 0)$.
For $x=0$ Eqs.~(\ref{eq:dyson2}), (\ref{eq:sigmad}), and (\ref{eq:2hedin}) correspond to a subset of the so-called Hedin's equations \cite{hedin65}, obtained by fixing $W$. A pictorial representation of this subset for a given $W$ is given in Fig.~\ref{fig:8}. In the following we will approximate the equations and the results will be compared to the exact solution of the differential equation, in order to obtain greater insight about these self-energy based techniques. From now on all quantities will hence be understood to be taken at $x=0$.
\subsubsection{$G_0W_0$ and self-consistency}
\begin{figure}
\includegraphics[width=3.6in,angle=360]{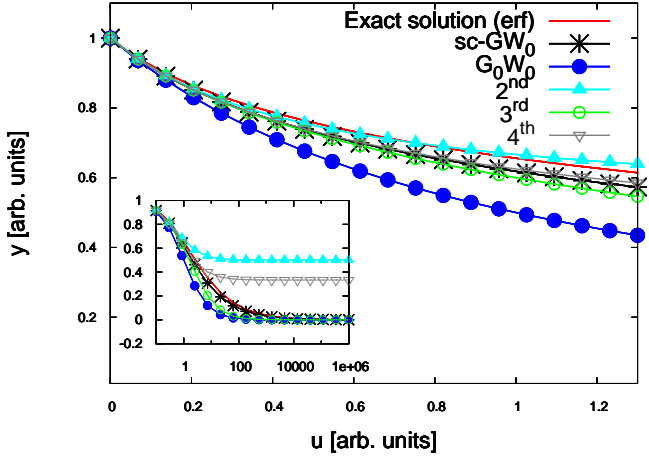}
\caption{Comparison between the exact solution (red plain line, Eq.\ (\ref{eq:gensolode})) and different flavours of the GW approximation. In general the self-energy based approximations perform  better than the iteration of the DE shown in Fig.\  \ref{fig:1}. In the main panel the self-consistent $GW_0$ (black stars, Eq.\ (\ref{eq:gwy})) is the best approximation to the exact result. Iterations starting from $G=G_0$ converge towards the self-consistent result (the $2^{nd}$ iteration is represented with light blue triangles, the $3^{rd}$ with green circles and the $4^{th}$ with grey empty triangles). However, analyzing a larger $u$ range (inset), one observes that odd iterations approach the exact $u=\infty$ limit, while the even ones don't seem to. It can be shown that they also do, however in a very slow fashion and according to the following sequence $y^{(2n)}_{u \rightarrow \infty} = \left\{ 1/2, 1/3, 1/4, 1/5, 1/6, \cdots \right\}$.}
\label{fig:3}
\end{figure}

Let us first look at different flavours of the $GW$ approximation \cite{hedin65}. 
Setting $\displaystyle \Gamma_u (y_u) $ to unity, it follows that $\displaystyle \Sigma_u(y_u) = -uy_u$. Within the initial guess $ y_u^{(0)} = y_0$, one obtains a so-called $G_0 W_0$ self-energy $ \Sigma_u=-uy_0$.\footnote{In realistic calculations $G_0$ is often taken to be a Kohn-Sham Green's function; here, to be consistent, it corresponds to the Hartree Green's function $G_H^0$.} This is then employed in the Dyson equation (\ref{eq:dyson2}) in order to get an improved  $y_u^{(1)}$. To go beyond this first approximation one can iterate further within the $GW$ approximation, i.e. keeping $\Gamma_u=1$. This corresponds to an iteration towards a $GW_0$ result, since $G$ is iterated towards self-consistency but $u$, which represents the screened interaction, is kept fixed. 
 We report here the expressions obtained for $G_0W_0$, i.e. the first solution of the Dyson equation, and for three successive loops 
\bea
\displaystyle
 \label{eq:gowo}
  y^{(1)}_u = y^{G_0 W_0}_u  &= & \frac{y_0}{1 + uy_0^{2}}, \\
  y^{(2)}_u &=& y_0\frac{1 + uy_0^{2}}{1 + 2u y_0^{2}} ,\\
  y^{(3)}_u &=& y_0 \frac{1 +2uy_0^{2} }{1 + 3uy_0^{2} + u^{2}y_0^{4}} , \\
  \label{eq:3rdloop}
  y^{(4)}_u &=& y_0 \frac{1 + 3u y_0^{2} + u^{2}y_0^{4}}{1 + 4uy_0^{2} + 3u^{2}y_0^{4}}.
\eea
We call this procedure \textit{iterative self-consistent} scheme, in contrast with the \textit{direct self-consistent} scheme where one solves directly the Dyson equation (\ref{Eqn:Dyson_eq}), for $x=0$, with $\Sigma_u=-uy_u$. In this latter case one gets a second-order equation with two solutions 
\bea
 \label{eq:gwy}
 \displaystyle
 y_u = \frac{\pm\sqrt{1 + 4u y_0^{2}}-1}{2 u y_0} .
\eea
Note that for the full functional problem one would find even more solutions, since a second-order equation has to be solved for each matrix element of $G$. \\
In order to choose the physical solution we Taylor expand the square root around $u=0$, which leads to
\bea
 y_u \approx \pm \left(y_0 + \frac{1}{2uy_0}\right) - \frac{1}{2uy_0}.
\eea
Since for $u=0$ one has to obtain $\displaystyle y_u=y_0$, the physical solution is $ \displaystyle y_u = \frac{\sqrt{1 + 4u y_0^{2}}-1}{2 u y_0}$. 
In Fig.~\ref{fig:3} we can appreciate how well these $GW$-based methods are performing against the exact solution in a wide $u$ range. \\
Interestingly \textit{odd} iterations quickly converge to the \textit{physical} solution of the direct sc-$GW_0$, while \textit{even} iterations do also converge but at a slower pace: it can be shown that for $u \rightarrow \infty$ their limit forms the sequence of rational numbers $\displaystyle \left\{ \frac{1}{2}; \frac{1}{3}; \frac{1}{4}; \frac{1}{5}; \frac{1}{6} \cdots \right\}$ which ultimately converges to 0. All the \textit{odd} iterations have instead the exact large $u$ limit (namely $y_u = 0$ when $u \rightarrow \infty$).  One might use this property to improve the convergence of the series. \\
An important question is now: \textit{does the result of the self-consistent procedure depend on the starting point of the iteration?}
Here we have naturally chosen $\displaystyle y_u^{(0)}=y_0$, but one might fear that this choice is simply lucky. Let us therefore look at the general iterative scheme which is obtained by solving the Dyson equation (\ref{Eqn:Dyson_eq}) for $x=0$ 
\begin{equation}
 y_u=\cfrac{y_0}{1+y_0uy_u}.
\end{equation}
By starting the iteration with a guess for $y_u$  on the right side one obtains 
\begin{equation}
 y_u^{(n+1)} = \cfrac{y_0}{1+y_0uy_u^{(n)}}. 
\end{equation}
For $y_u^{(0)}=y^{s}$ one has e. g. after the third iteration
\begin{equation}
y^{(3)}_u = \cfrac{y_0}{1+\cfrac{uy_0^2}{1+\frac{uy_0^2}{1+y_0uy^s}}}.
\end{equation}
This contains nothing else but the continued fraction representation for the square root
\begin{equation}
\sqrt{1+z}=1+\cfrac{z/2}{1+\cfrac{z/4}{1+\cfrac{z/4}{1+\cfrac{z/4}{1+\cfrac{z/4}{1...}}}}},
\label{eq:contfrac}
\end{equation}
corresponding to the physical solution $\displaystyle y_u = \frac{\sqrt{1+z}-1}{2uy_0}$ where $\displaystyle z=4uy_0^2$. It converges for all values of the terminator $\displaystyle y^s$. Therefore, \textit{this iteration will always converge to the physical solution.}
Does this mean that there is no risk of running into the unphysical solution? 
The answer is that it depends on the iterative scheme that is used, and \textit{not} on the starting point. Look at the following way to re-write the Dyson equation (\ref{Eqn:Dyson_eq}):  $\displaystyle - uy_u = \frac{1}{y_0} - \frac{1}{y_u}$ (in other words, $\Sigma = G_0^{-1}-G^{-1} $). If we iterate this equation by starting with some $y_u^{(0)} = y^s$ on the right-hand side we get
\begin{equation}
 y_u^{(n+1)} = -\frac{1}{uy_0}+\frac{1}{u y_u^{(n)}},
\end{equation}
hence
\begin{equation}
2uy_0y = - 2 - \cfrac{2uy_0^2}{1+\frac{uy_0^2}{1+\frac{uy_0^2}{1+\frac{uy_0^2}{1+\frac{uy_0^2}{\dots}}}}}.
\end{equation}

which, with Eq.~(\ref{eq:contfrac}), is just the continued fraction representation for the unphysical solution $y_u=(-\sqrt{1+4uy_0^2}-1)/2uy_0 $.  
In a way, this is good news: usually the iterative scheme adopted in the context of $GW$ calculations is rather the first, safe one. 
Indeed it has been found empirically that such a scheme leads to self-consistent results independent of the starting point and in reasonably good agreement with experiments (see e.g.~\cite{stan-2006, stan-2009, marsili}).
However, when one goes beyond GW, higher order equations appear as we will see in the following. There are hence more and more solutions, and more and more ways to iterate the equations. In other words, \textit{there will be an increased danger to run into a wrong solution}. One should keep this in mind when trying to add vertex corrections beyond $GW$. 
%
%
%
%
\subsubsection{Vertex corrections - First order $\Gamma$}

We will now analyze the effects of a first order vertex correction which is obtained employing  $\displaystyle \Sigma_u=-uy_u$ in Eq.~(\ref{eq:2hedin}) ~\cite{hedin65,shirley}. Solving for $\Gamma_u$ gives
\bea
 \displaystyle
 \label{eq:gamma1}
 \Gamma_u^{(1)}(y_u) = \frac{1}{1 + u y_u^{2}}.
\eea
Employing this vertex the self-energy (\ref{eq:sigmad}) becomes
\begin{eqnarray}
 \label{eq:self-cor}
\displaystyle
\Sigma_u^{(1)} (y_u) = -uy_u \ob \frac{1}{1 + uy_u^{2}} \cb,
\end{eqnarray}
Now two routes can be taken and either a $G_0 W_0\Gamma^{(1)}(y_0)$ or a self-consistent $GW_0\Gamma^{(1)}(y_u)$ calculation can be carried out. The first of the two is once more based on the initial guess for the Green's function $y^{(0)} = y_0$, and consequently the vertex and the self-energy in (~\ref{eq:gamma1}-\ref{eq:self-cor}) read respectively $\displaystyle \Gamma_u^{(1)}(y_0) = \frac{1}{1 + u y_0^{2}}$ and $\displaystyle \Sigma_u^{(1)}(y_0)= -uy_0 \ob \frac{1}{1 + uy_0^{2}} \cb$. Solving the Dyson equation with the above ingredients yields:
\begin{equation}
 \displaystyle
\label{Eqn:g0w0gamma}
y^{G_0 W_0\Gamma}_u = \frac{y_0 \left( 1 + uy_0^{2}  \right)}{1 + 2uy_0^{2}}. 
\end{equation}

\begin{figure}
\includegraphics[width=3.6in,angle=360]{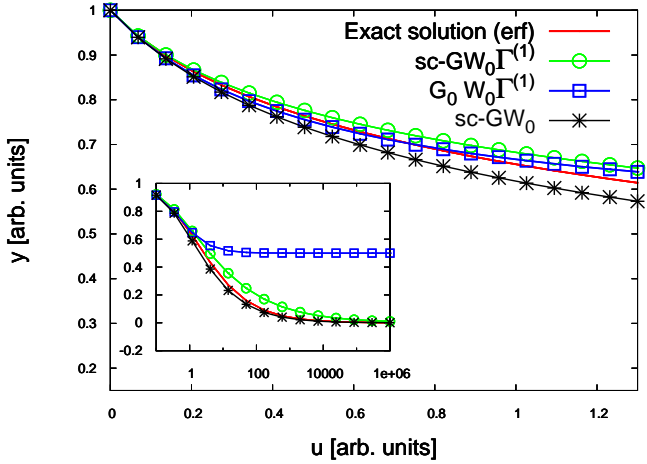}
\caption{In the main panel a comparison between the DE's exact solution (red plain line, Eq.\ (\ref{eq:gensolode})), $G_0W_0 \Gamma^{(1)}$ (blue squares, Eq.\ (\ref{Eqn:g0w0gamma})), $GW_0 \Gamma^{(1)}$ (green empty circles, Eq.\ (\ref{eq:ygwgam})) and sc-$GW_0$ (black stars, Eq.\ (\ref{eq:gwy})) is shown. In this range of $u$, adding a vertex correction, no matter if within a self-consistent scheme or not, improves over the simpler self-consistent $GW_0$. However, analyzing a wide $u$ range (inset, semi-logarithmic plot), gives a different perspective: the first iteration of $G_0W_0 \Gamma^{(1)}$ clearly exhibits the wrong $u \rightarrow \infty$ limit and the sc-$GW_0$ scheme becomes the closest approximation to the exact result.}
\label{fig:4}
\end{figure}

Instead, solving the Dyson equation in a self-consistent fashion, with the expressions (\ref{eq:gamma1}-\ref{eq:self-cor}) yields:
\begin{eqnarray}
 \label{eq:ygwgam}
\displaystyle
y^{GW_0\Gamma}_u & =&  \sqrt[3]{\frac{y_0}{2u}  + \sqrt{ \frac{1}{27 u^{3}} + \frac{1}{4 u^{2}}}} \nonumber \\
&-& \sqrt[3]{\frac{y_0}{2u} - \sqrt{ \frac{1}{27 u^{3}} + \frac{1}{4 u^{2}}}}.
\end{eqnarray}
As it can be noticed from the result a cubic equation for the unknown $y_u$ had to be solved within this more sophisticated approach. Again the limit of vanishing interaction has been used to pick the physical solution.
In Fig.~\ref{fig:4} we can directly compare the two types of vertex corrections. For small $u$ values their performance is similar, however, in a wider $u$ range (see inset), the $G_0 W_0\Gamma^{(1)}$ scheme diverges from the exact solution and has the wrong asymptotic limit $u\rightarrow \infty$: it hence behaves as the first iteration of the sc-$GW_0$ approach, which also exhibits the wrong large $u$ limit.
Fig.~\ref{fig:4} also shows how the $GW_0\Gamma^{(1)}$ scheme, for small $u$ values, slightly improves over the sc-$GW_0$. However, given the augmented complexity already at this first order of the correction  (one could very well iterate further the equations for $\Gamma$ and $\Sigma$ and get higher order corrections), the benefits of employing vertex corrections are not obvious. Also note that interestingly, on the scale from $u=0$ to $u \rightarrow \infty$, the closest curve to the exact one is the sc-$GW_0$ one.
%
%
%
\section{Exploring other approximations for G}
\label{sec:newmethod}
In this section we will explore alternative approximations to the exact solution of the 1-point DE  and the corresponding manipulations of the initial differential equation producing them.
Here we will report in particular approximations that might be eventually transposed to the full functional framework. 
%
%
%
%
\subsection{Continued fraction approximation}
\begin{figure}
\includegraphics[width=3.4in,angle=360]{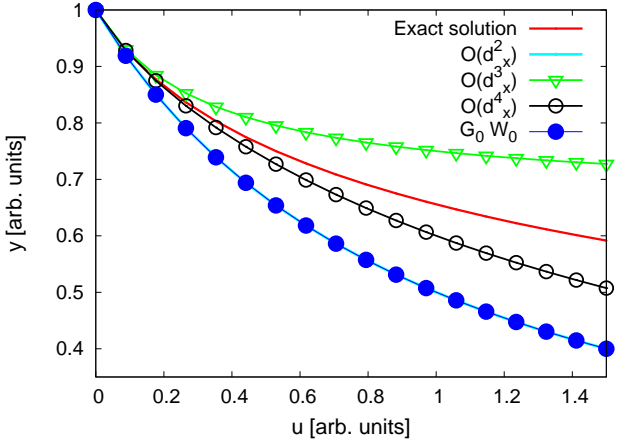}
\caption{Comparison between the exact solution (red plain line, Eq.\
(\ref{eq:gensolode})) of the DE and the results obtained through the
first three orders of the \textit{limited order differential equation}
(Eq.\ (\ref{eq:trunres0})). The notation $O(d^{n}_x)$
indicates that derivatives of order $\geq n$ have been neglected. As
expected the result improves when more terms are included: the curve
$\displaystyle O(d^{2}_x)$ (light blue line,  Eq.\
(\ref{eq:tay})) is superimposed to the $G_0W_0$ one (dark blue dots,
Eq.\ (\ref{eq:gowo})) and the curve $\displaystyle
O(d^{4}_x)$ (black circles,  Eq.\ (\ref{eq:taytru})) is
close to the exact result in a small $u$ range.}
\label{fig:2}
\end{figure}

A well known approximation for the error function is its \textit{continued fraction} representation ~\cite{abramowitz}. The exact expression for $y_u$ (Eq.~(\ref{eq:gensolode})) transforms into
\bea
 \displaystyle
 \label{eq:contf1}
 && y_u = \cfrac{1}{\sqrt{2u}}  \nonumber \\
 &&\times \cfrac{1}{\cfrac{1}{\sqrt{2uy_0^{2}}}+\cfrac{1/2}{\cfrac{1}{\sqrt{2uy_0^{2}}}+\cfrac{1}{\cfrac{1}{\sqrt{2uy_0^{2}}}+\cfrac{3/2}{\cfrac{1}{\sqrt{2uy_0^{2}}}+\dots}}}}  \\
 \label{eq:contf2}
 && =  \cfrac{y_0}{1 + \cfrac{uy_0^{2}}{1 + \cfrac{2uy_0^{2}}{1+ \cfrac{3uy_0^{2}}{1 + \dots} }}}. 
\eea
We will now show how one can obtain Eq.\ (\ref{eq:contf2}) starting simply from the initial DE in Eq.~(\ref{Eqn:diff0dnotr}), equivalent to (\ref{eq:nohart}), without any information about its exact solution. 
Beginning with Eq.~(\ref{Eqn:diff0dnotr}) and taking successively higher order derivatives of the equation, one obtains :
\begin{eqnarray}
 \label{eq:tay}
\displaystyle
&&\frac{d y_u(x)}{d x} = \! y_0 y_u(x) + y_0x \frac{d y_u(x)}{d x} -u y_0 \frac{d^{2} y_u(x)}{d x^{2}} \\
\label{eq:taytru1}
&&\frac{d^{2} y_u(x)}{d x^{2}} = \! 2 y_0 \frac{d y_u(x)}{d x} \! + y_0 x \frac{d^{2} y_u(x)}{d x^{2}}   -u y_0 \frac{d^{3} y_u(x)}{d x^{3}} \\
\label{eq:taytru}
&& \frac{d^{3} y_u(x)}{d x^{3}} = \! 3 y_0 \frac{d^{2} y_u(x)}{d x^{2}}  + \! y_0x \frac{d^{3} y_u(x)}{d x^{3}}  - uy_0 \frac{d^{4} y_u(x)}{d x^{4}}
\end{eqnarray}
and so on. Neglecting derivatives e.g. from the $4^{th}$ order on and then setting $x=0$, this \textit{truncation} allows us to solve all the above equations, beginning with Eq.~(\ref{eq:taytru}) (now an algebraic equation in the unknown $\displaystyle \frac{d^{3} y_u(x)}{d x^{3}}$ by keeping $\displaystyle \frac{d^{2} y_u(x)}{d x^{2}}$ as parameter); subsequently we insert the result in (\ref{eq:taytru1}) and solve for $\displaystyle \frac{d^{2} y_u(x)}{d x^{2}}$, (\ref{eq:tay}) for $\displaystyle \frac{d y_u(x)}{d x}$ and ultimately Eq.~(\ref{Eqn:diff0dnotr}) getting
\begin{eqnarray}
 \displaystyle
 \label{eq:trunres0}
 y_u= \cfrac{y_0}{1+\cfrac{uy_0^{2}}{1+\cfrac{2uy_0^{2}}{1+ 3uy_0^{2}}}},
\end{eqnarray}
which is precisely the result obtained by approximating the exact solution with a continued fraction expression for the error function ( Eq.~(\ref{eq:contf2}) ). We will name this manipulation \textit{limited order differential equation}. 
In Fig.~\ref{fig:2} we compare the different orders of this approximation to the exact expression for $y_u$. The approximation gets rapidly closer and closer to the exact solution by including higher derivatives.
However, also for this continued fraction, odd and even orders, converge towards the exact result with a different speed. In analogy with the continued fraction of Eq.~(\ref{eq:contfrac}), even iterations have the correct large $u$ limit, while the odd ones don't, although they do eventually approach it for a very large number of steps. We notice that the above continued fraction converges slower than the one arising from the sc-$GW_0$; however, the former will eventually converge towards the exact solution, whereas the latter only to the self-consistent $GW_0$ result.
It is therefore interesting to note that such a procedure can in principle be used also in the full functional framework (see related manipulations e.g. in~\cite{tay-yar,Csanak}), where the functional differential equation can be differentiated to an arbitrary order and the corresponding approximated $G$ obtained. For example, differentiating Eq.~(\ref{Eqn:linerized_ODE}) with respect to the external potential $\bar{\ep}$ one gets 
\bea
\displaystyle
\label{Eqn:cf1}
 \frac{\delta G(1,2;[\bar{\varphi}])}{\delta \bar{\varphi}(6)} & = & \int d3 G_H^0(1,3) \frac{\delta \bar{\varphi}(3)}{\delta \bar{\ep}(6)} G(3,2;[\bar{\varphi}]) \nonumber \\
 &+& \int d3 G_H^0(1,3) \bar{\ep}(3) \frac{\delta G(3,2;[\bar{\varphi}])}{\delta \bar{\ep}(6)} \nonumber \\
 & + & i \int d3 d5 W(3^{+},5) G_H^0(1,3) \nonumber \\ 
& \times & \frac{\dk^{2} G(3,2;[\varphi])}{\delta \bar{\ep}(6) \delta \bar{\ep}(5)}.
\eea
Truncating the highest order derivative $\displaystyle \frac{\dk^{2}G}{\dk \bar{\ep}^{2}}$ and solving for $\varphi = 0$ (which means also $\displaystyle \bar{\ep} = 0$) gives:

\bea
\displaystyle
\label{Eqn:cf1.1}
\left.\frac{\dk G(1,2;[\bar{\varphi}])}{\dk \bar{\ep}(5)} \right|_{\bar{\varphi}\rightarrow 0}= G_H^0(1,5)G(5,2) 
\eea
which reinserted in Eq.~(\ref{Eqn:linerized_ODE}) yields:
\bea
\displaystyle
\label{Eqn:cf1.2}
 G(1,2) &=& G_H^0(1,2) \nonumber \\ 
&+& i \int d3 d5 G_H^0(1,3)  W(3^{+},5) \nonumber \\ 
&\times& G_H^0(3,5) G(5,2).
\eea
%
Like in the 1-point model, this first step simply provides the one-particle GF in the $G^0_HW_0$ approximation to the self-energy. 
One can go further: differentiating Eq.\ (\ref{Eqn:cf1}) with respect to $\bar{\ep}$ and neglecting the third order derivative $\displaystyle \frac{\delta^3 G}{\delta \bar{\varphi}^3}$ yields:

\bea
\displaystyle
\label{Eqn:cfa11}
 G(1,2)  & = & G_H^0(1,2) -i \int d 5 d3 d8 d9 G_H^0(1,3) W(3^{+},5) \nonumber \\
&\times& \bar{m}^{-1}(3,5;9,8) G_H^0(9,8)G(8,2)       
\eea
with
\bea
\displaystyle
\label{Eqn:m}
 \bar{m}(16;57) :&=&-\delta(15)\delta(76)\nonumber\\
 & + & i\int d3  G_H^0(1,3) W(3^{+},5) \delta(7,6) \nonumber \\
 & \times& \left[ G_H^0(3,6) + G_H^0(3,5) \right], 
\eea
which is a four-point quantity of a similar complexity as the Bethe-Salpeter equation ~\cite{Strinati}. Indeed in the full functional problem the equations become quite involved since terms like $uy_0^2$ correspond to large matrices. However the approach doesn't require self-consistency. This might turn out to be a significant advantage, compared to vertex corrections to $\Sigma$, as we have discussed in the previous subsection concerning self-consistency. More details about the derivation are given in App.~\ref{sec:appcf}.
%
%
%
\subsection{Large $u$ expansions}
Perturbation theory usually deals with weak interactions, hence the small $u$ limit. However, it is also very interesting to examine the \textit{large u} limit for several reasons: \textbf{i}) this is the regime of  \textit{strong correlation}, where current approximations exhibit failures; \textbf{ii}) the large $u$ expansion of the exact solution gives a convergent series (being a product of two convergent Taylor expansions, one for the exponential and the other one for the error function) and one can, for instance, obtain a better approximation to the exact solution by adding higher order terms (which instead does not improve the result for the small $u$  expansion of the solution); \textbf{iii}) excellent approximations for the exact solution are \textit{Pad\'e approximants}~\cite{roy}, which have to be constructed using both the small and large $u$ limit.  
In this subsection we will present two possible routes to approach this limit: the first is a straightforward large $u$ expansion of the exact solution for $y_u$, while the second combines the latter with the large $u$ expansion for the Dyson equation.
\subsubsection{Straightforward large $u$ expansion for $y_u$ }
\begin{figure}
\includegraphics[width=3.4in,angle=360]{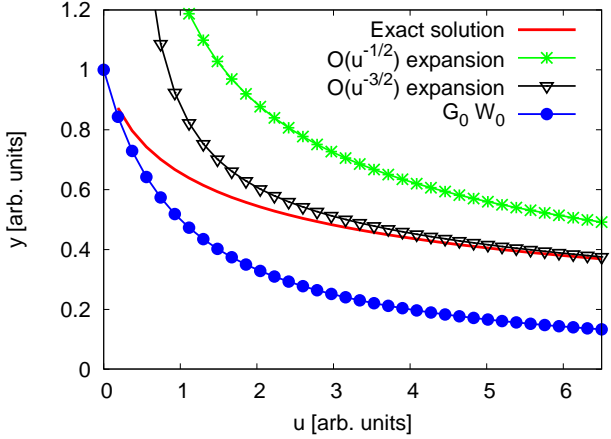}
 \caption{Comparison between the exact solution (red plain line, Eq.\ (\ref{eq:gensolode})) and the large $u$ expansion for the DE. The green stars and the black triangles are respectively $O(u^{-1/2})$ and $O(u^{-3/2})$ of the large $u$ expansion (Eqs.\ (\ref{eq:1}-\ref{eq:ouhigh})). We also report the $G_0W_0$ result (blue dots, Eq.\ (\ref{eq:gowo})) as an example of a small $u$ expansion. Over a wide $u$ range the large $u$ expansions are very satisfactory.}
\label{fig:6}
\end{figure}

By expanding both the exponential prefactor and the error function appearing in Eq.\ (\ref{eq:gensolode}):
\begin{eqnarray}
 \label{eq:largeu2terms}
 e^{\frac{1}{2uy_0^{2}}}& \approx &1 + \frac{1}{2uy_0^{2}} + \frac{1}{8u^{2}y_0^{4}} + \cdots, \\
erf \left[ \sqrt{\frac{1}{2uy_0^{2}}} \right] &\approx& \frac {2}{\sqrt{\pi}} \ob \sqrt{\frac{1}{2uy_0^{2}}} - \frac{1}{6uy_0^{2}} \sqrt{\frac{1}{2uy_0^{2}}} \nonumber \\
&+& \frac{1}{40 u^{2} y_0^{5}} \sqrt{\frac{1}{2uy_0^{2}}} + \cdots \cb,
\end{eqnarray}
 one obtains for the different orders of the full solution
\begin{eqnarray}
 \label{eq:largeuexp}
y^{(-1/2)}_u &=& \sqrt{\frac{\pi}{2u}}\label{eq:1} \\
\label{eq:ou}
y^{(-1)}_u &=& -\frac{1}{uy_0 } + \sqrt{\frac{\pi}{2u}} \\
\label{eq:ouhigh}
y^{(-3/2)}_u &=& -\frac{1}{uy_0 } +\frac{1}{2uy_0^{2}} \sqrt{\frac{\pi}{2u}} + \sqrt{\frac{\pi}{2u}} \\
 y^{(-2)}_u &=& -\frac{1}{uy_0 } +\frac{1}{2uy_0^{2} }\sqrt{\frac{\pi}{2u}} -\frac{1}{6u^{2}y_0^{3}}+ \sqrt{\frac{\pi}{2u}} \\
 y^{(-5/2)}_u &=& -\frac{1}{uy_0 } +\frac{1}{2uy_0^{2} }\sqrt{\frac{\pi}{2u}} -\frac{1}{6u^{2}y_0^{3}} \nonumber \\
 &+& \frac{1}{8u^{2}y_0^{4}} \sqrt{\frac{\pi}{2u}} + \sqrt{\frac{\pi}{2u}} \\
 y^{(-3)}_u &=& -\frac{1}{uy_0 } +\frac{1}{2uy_0^{2} }\sqrt{\frac{\pi}{2u}} -\frac{1}{6u^{2}y_0^{3}} +  \frac{1}{8u^{2}y_0^{4}}
 \sqrt{\frac{\pi}{2u}} \nonumber \\
 &+& \frac{1}{10u^{3}y_0^{5}} + \sqrt{\frac{\pi}{2u}}. 
\end{eqnarray}
Fig.~\ref{fig:6} shows how these different expansions perform \textit{versus} the exact result. Overall their behaviour is very good for large $u$ and  few orders are sufficient to get a good approximation over a wide $u$ range (which is our ultimate goal), however for $u=0$ they all diverge.
%
%
%
%
%
\subsubsection{Large $u$ expansion for $y_u$ and for the Dyson equation }
\begin{figure}
\includegraphics[width=3.4in,angle=360]{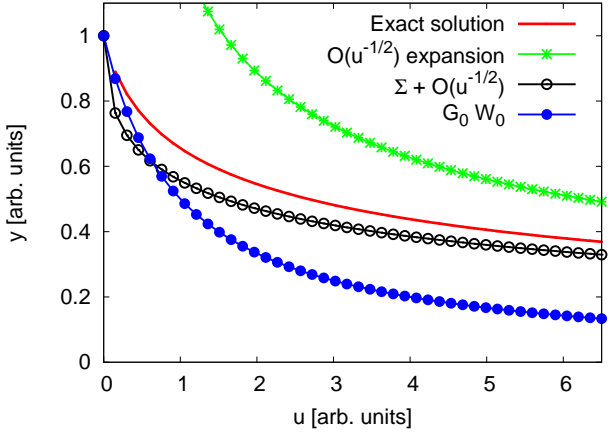}
\caption{Comparison between the exact solution (plain red line, Eq.\ (\ref{eq:gensolode})), the $G_0W_0$ result, the order $O(u^{-1/2})$ for the DE's large $u$ expansion (green stars, Eq.\ (\ref{eq:1}) ), and the same order of the DE's large $u$ expansion combined with the large $\Sigma$ expansion (black dots, Eq.\ (\ref{eq:dysexp5})). We observe that the latter approximation performs extremely well over the range of interaction examined, being even exact both in the large and small $u$ limits.}
\label{fig:7}
\end{figure}

When $u$ gets larger, also $\Sigma_u$ increases. This implies that, using the Dyson equation for the one-particle Green's function $y_u =\left( y_0^{-1} - \Sigma_{u}\right)^{-1}$ one could expand $y_u$ as
\begin{eqnarray}
 \label{eq:dysexp1}
 y_u  &\approx& - \Sigma_u^{-1} \ob 1 + y_0^{-1} \Sigma_u^{-1} +  y_0^{-1} \Sigma_u^{-1} y_0^{-1} \Sigma_u^{-1} \cb.
\end{eqnarray}
Hence to lowest order $\displaystyle y_u \approx - \Sigma_u^{-1}$ or
\begin{eqnarray}
 \label{eq:dysexp2}
 \Sigma_u \approx -\frac{1}{y_u}.
\end{eqnarray}
This simple relation allows us to use the large $u$ expansion of the exact solution for $y_u$ to approximate $\Sigma_u$ for large $u$; we can then use this approximate $\displaystyle \Sigma_u$ in the Dyson equation to recalculate $y_u$. For example, using the lowest order of the large $u$ expansion of the exact $y_u$ one gets the following self-energy:
\begin{eqnarray}
 \displaystyle
  \label{eq:dysexp4}
 \Sigma_u \approx -\left(\sqrt{\frac{\pi}{2u}} \right)^{-1},
\end{eqnarray}
which reinserted in the Dyson equation $\displaystyle y_u=(y_0^{-1}-\Sigma_u)^{-1}$ gives:
\begin{eqnarray}
\displaystyle
  \label{eq:dysexp5}
  y_u  &\approx& \cfrac{y_0}{1+ y_0\sqrt{\frac{2u}{\pi}}}.
\end{eqnarray}
In Fig.~\ref{fig:7} the performance of this approximation for $y_u$ is plotted against two orders of the straightforward large $u$ expansion for the Green's function, $G_0W_0$ and the exact solution. 
The "large $\Sigma$" approach shows an overall good agreement (generally better than $G_0W_0$) with the exact solution and it has the desirable property of being exact in the small and large $u$ limits, mending the divergence of all orders of the straightforward expansion for $y_u$. 
At higher orders of the approximation this property remains true, although undesired poles appear.
In conclusion the methodology is promising and worthwhile to be explored further. The main difficulty is that in the framework of a large $u$ expansion, without knowing the exact solution, one would not straightforwardly know how to set the constant $C$, i.e. how to pick the physical solution: this issue requires further analysis.
%
%
%
\subsection{Self-consistent calculations of the Hartree Green's function and of the screened interaction}
In the above discussions we have treated the Hartree Green's function and the screened interaction as externally given quantities. 
This is justified by the fact that realistic calculations are most often following such a pragmatic ansatz. In principle these quantities should be part of Hedin's self-consistent cycle. A fully self-consistent treatment, in the full functional framework, is today out of reach. In the 1-point model, however, it is possible to go beyond this limitation and indeed, the implicit solution of Hedin's equation that has been achieved in the work of~\cite{hubner} contains all quantities calculated on the same footing. Also in the linearized version, that is employed in the present work, one can obtain the Hartree Green's function and the screened potential consistently from the equations, as we will discuss in the following.
Let us first turn to the Hartree Green's function $y_H^0$. In terms of the truly non-interacting Green's function $y_0$ it reads
\begin{equation}
 y_H^0 = \frac{y_0}{1-y_0uy_u},
\label{eq:sc-h}
\end{equation}
in other words, it depends (through the density) on the solution $y_u$ at vanishing external potential. In a self-consistent scheme this $y_H^0$ should then replace $y_0$ in the solution Eq.~(\ref{eq:gensolode}), which leads to an implicit equation for $y_u$. 
For a self-consistent treatment of the screened interaction we can use of the fact that the 1-point differential equation can be solved for $\displaystyle \frac{d y_u}{dx}$, and insert the result into the expression for the screened interaction $u$ in terms of the bare $v$, which reads $\displaystyle u = v + v \frac{dy_u}{dx} v $. Two routes can be taken. The first one is based on the linearized equation  
(\ref{Eqn:diff0dnotr}) where the interaction is already screened from the very beginning. This leads to a quadratic equation for $u$, with two solutions
\begin{equation}
\displaystyle
 u =  \frac{v}{2} \pm \sqrt{ \frac{v^2}{4} + v^2 \left\{ 1 - \frac{y_u}{y_0} + vy^2_u \right\} }   .
\label{eq:sc-w-1}
\end{equation}
The physical solution is the one of the positive square root, since it approaches the bare $v$ in the limit of vanishing interaction, hence vanishing screening. The second route consists in calculating $\displaystyle \frac{dy_u}{dx}$ from the initial equation (\ref{Eqn:diff0dfull}), where the bare $y_0$ and the interaction $v$ appear. This yields
\begin{equation}
 u = v \left( 2-\frac{y_u}{y_0} + v y_u^2 \right).
\label{eq:sc-w-2}
\end{equation}
In both cases, the solution for $u$ should be used into Eq.~(\ref{eq:gensolode}), which again makes the expression for the GF implicit. 
One may argue about which of the two ways to calculate $u$ self-consistently is more adequate. 
In a realistic calculation one would probably use the former approach, in an iterative way: after calculating the GF as a functional of the external potential for a given initial interaction in the linearized DE, one would recalculate the $W$ from the functional derivative, and so on. Whatever choice, however, does not influence the main conclusions that can be drawn from the above considerations. Specifically: \textbf{i)} a self-consistent calculation leads to an implicit solution (like in the work of~\cite{hubner}) which however would not be identical to theirs because of our linearization procedure; \textbf{ii)} the behaviour for the small interaction limit is unchanged by the self-consistent treatment, as one can verify from equations (\ref{eq:sc-h}),(\ref{eq:sc-w-1}) and (\ref{eq:sc-w-2}); this means in particular that the constant $C$ is chosen in the same way as before. \textbf{iii)} Finally also the discussion about the limit of large interaction is unchanged: by making the ansatz that to lowest order $\displaystyle y_u \propto \frac{1}{\sqrt{u}}$ one finds consistency. \\
Altogether, this shows that the linearization of the equations does not imply necessarily that one has to treat the Hartree Green's function and the screened interaction as externally given quantities. It also shows that a more refined, self-consistent treatment does not change the overall behaviour of the solution.
%
%
\section{Conclusions and outlook}
\label{sec:concl}
In this paper we explore several aspects of the set of first order nonlinear coupled differential equations which are conventionally solved perturbatively in order to calculate the one-particle Green's function.
After the linearization of the Hartree potential with respect to the external one, we employ a 1-point model where the set of -now linear- differential equations reduces to a $1^{st}$ order algebraic differential equation, that can be solved exactly. This provides insights into the structure of the general family of solutions, and on how to determine which amongst them corresponds to the physical one. 
Within the model we study the performance of established approaches over the whole range of interaction strength: we find that iterations towards self-consistency in the $GW$ scheme sensibly improve on the one-shot ($G_0W_0$) calculation and that including first order vertex corrections improves the self-consistent $GW_0$ results only slightly and only for small $u$. We also find that in case of self-consistent $GW_0$ two solutions are possible, of which only one is physical.
We show that the standard iterative scheme will always converge to the physical solution, although other schemes may yield different results. 
This is an important finding: when going beyond $GW$ both the number of possible solutions for the Green's function and the number of possible ways to iterate the equations increase, creating a danger to run into a wrong solution. 
We finally explore other approximations to the exact solution that might be transposed to the full functional framework, namely a continued fraction approximation and the expansion for large interaction, and we relate these approximations to the corresponding manipulations of the differential equation which produce them. These links are crucial to prepare a generalization of the approach to the full functional framework.
\begin{acknowledgments}
We would like to acknowledge fruitful discussions with C.~Brouder and R.~W. Godby. The work was supported by ANR (Project No. NT09-610745 ). 
\end{acknowledgments}
\appendix
%
%
\section{Approximation for the Hartree term \label{sec:appahar}}
Due to the Hartree potential $V_H=-ivG$ the set of differential equations (\ref{eq:diffnd}) is nonlinear. In order to simplify this problem we first assume that $V_H$ is Taylor expandable in terms of the external potential $\varphi$:
\bea
 \displaystyle
 V_H(3;[\varphi]) &\approx& - i \int d4 v(3^{+},4) G(4,4^{+};[\varphi]) \Big{|}_{\varphi = 0} \nonumber \\
 &-& i\int d4 d5 v(3^{+},4) \left.\frac{\delta G(4,4^{+};[\varphi])}{\delta \varphi(5)}\right|_{\varphi = 0} \varphi(5) \nonumber \\
 &+& o(\varphi ^2).
\eea
The second step is to introduce $G^{0}_H$ defined through 
\bea
 \displaystyle
 G^{0}_H(1,2) &=&  G_0(1,2) \nonumber \\ 
 &+& \int d3 G_0(1,3) V^{0}_H(3) G^{0}_H(3,2),
\eea
with $\displaystyle V^{0}_H(3) := -i \int d4 v(3^{+},4) G(4,4^{+};[\varphi]) \Big{|}_{\varphi = 0}$.
Inserting $V_H$ in Eq. (\ref{eq:diffnd}) one obtains
\bea
 \displaystyle
\label{eq:appahar1}
 && G(1,2;[\varphi]) = G^{0}_H(1,2) +  \int d3 d5 G^{0}_H(1,3) \nonumber \\
 && \times \ob -i\int d4 d5 v(3^{+},4)  \frac{\delta G(4,4^{+};[\varphi]) }{\delta \varphi(5)}\Big{|}_{\varphi = 0} \nonumber \\
 && \, + \delta(3,5) \cb \: \varphi(5) G(3,2;[\varphi]) \nonumber \\
 && \, + i \int d3 d4 G^{0}_H(1,3) v(3^{+},4) \frac{\delta G(3,2;[\varphi])}{\delta \varphi(4)}
\eea
Since $\displaystyle \frac{\delta G}{\delta \varphi}$ in the second term on the right-hand side of Eq.~(\ref{eq:appahar1}) is a contraction of the two-particle correlation function, it yields the inverse dielectric function
\begin{equation}
 -i\int d4 v(3^{+},4)  \frac{\delta G(4,4^{+};[\varphi])}{\delta \varphi(5)}  \Big{|}_{\varphi = 0} + \delta(3,5) = \epsilon^{-1}(3,5),
\end{equation}
and one gets
\bea
 \displaystyle
 \label{eq:appahar2}
 && G(1,2;[\varphi]) = G^{0}_H(1,2) +  \int d3 d5 G^{0}_H(1,3) \nonumber \\
 && \times \epsilon^{-1}(3,5) \varphi(5) G(3,2;[\varphi]) + i \int d3 d4 G^{0}_H(1,3) \nonumber \\
 && \times v(3^{+},4) \frac{\delta G(3,2;[\varphi])}{\delta \varphi(4)}.
\eea
Now a \textit{rescaled} perturbing potential can be introduced:
\begin{eqnarray}
 \displaystyle
\bar{\varphi}(3) := \int d5  \epsilon^{-1}(3,5) \varphi(5),
\end{eqnarray}
and, using the chain rule $\displaystyle \frac{\delta G}{\delta \varphi} 
= \frac{\delta G}{\delta \bar{\varphi}} \frac{ \delta \bar{\varphi}}{\delta \varphi}$ in the last term of the right-hand side of Eq.~(\ref{eq:appahar2}), we get:
\bea
 \displaystyle
\label{eq:appahar3}
 && G(1,2;[\bar{\varphi}]) = G^{0}_H(1,2) + \int d3 d5 G^{0}_H(1,3) \nonumber \\
 && \times \bar{\varphi}(3) G(3,2;[\bar{\varphi}]) + i \int d3 d5 G^{0}_H(1,3) \nonumber \\
 && \times W(3^+,5) \frac{\delta G(3,2;[\bar{\varphi}])}{\delta \bar{\varphi}(5)}, 
\eea
which is precisely Eq.~(\ref{eq:nohart}). Here $W=\epsilon^{-1}v$ is the screened Coulomb potential at vanishing $\varphi$. 
If one approximates the functional derivative $\displaystyle \frac{\delta G}{\delta \bar{\varphi}}=\displaystyle -G \frac{\delta G^{-1}}{\delta\bar{ \varphi}} G\approx GG$ which comes from assuming the self-energy in the Dyson equation $G^{-1} = G_0^{-1} -v_H^0- \Sigma - \bar{\varphi}$ to be independent of ${\bar{\varphi}}$, Eq.~(\ref{eq:appahar3}) becomes:
\bea
 \displaystyle
\label{eq:appahardy}
 && G(1,2;[\bar{\varphi}]) = G^{0}_H(1,2) + \int d3 d5 G^{0}_H(1,3) \nonumber \\
 && \times \bar{\varphi}(3) G(3,2;[\bar{\varphi}]) +  \int d3 d5 G^{0}_H(1,3) \nonumber \\
 && \times \Sigma_{GW}(3,5;[\bar{\varphi}]) G(5,2;[\bar{\varphi}])
\eea
with $\displaystyle \Sigma_{GW}(3,5;[\bar{\varphi}])=iG(3,5; [\bar{\varphi}])W(3^+,5)$. For $\varphi=0$ Eq.~(\ref{eq:appahardy}) becomes the Dyson equation for the one-particle Green's function in the GW approximation to the self-energy \footnote{To be precise, here it is not specified how $W$ is obtained, it is in principle the exact $W$, whereas in $GW$ the screened interaction is usually calculated in the random phase approximation (RPA) }.
This confirms that the linearization of $V_H$ is a reasonable starting point for further developments.%
%
%
%
%
\section{Solving the DE\label{sec:appab}}
Eq.~(\ref{Eqn:diff0dnotr}) can be solved using standard textbook methods \cite{kamke, bronshtein}. Here we choose a route that yields precious information for our final aim of generalizing to the full functional problem.
A general ansatz for the structure of $y_u(x)$ is:
\begin{eqnarray}
 \displaystyle
y_u(x) = A(x) \cdot \mathcal{I}(x),
\end{eqnarray}
where the only restriction is that $A$ and $\mathcal{I}$ are not zero. Substituting the ansatz in the DE (\ref{Eqn:diff0dnotr}) gives:
\bea
 \displaystyle
 \label{appab:fac2}
 A(x) \mathcal{I}(x) &=& y_0 + y_0 x A(x)\mathcal{I}(x) - uy_0 \frac{d A(x)}{d x} \mathcal{I}(x) \nonumber \\
 &-& u y_0 A(x) \frac{d \mathcal{I}(x)}{d x}. 
\eea
The idea is now to solve two separate, simpler with respect to the initial one, DEs for $A(x)$ and $\mathcal{I}(x)$. Putting together the left-hand side and the second and third terms of the right-hand side of Eq.~(\ref{appab:fac2}) one obtains:
\begin{eqnarray}
 \displaystyle
 \label{appab:eq1}
 A(x) \mathcal{I}(x) = y_0x A(x) \mathcal{I}(x) -  uy_0 \frac{d A(x)}{d x} \mathcal{I}(x).
\end{eqnarray}
We can choose the solution
\begin{eqnarray}
  \displaystyle
  A(x) =  e^{ \left[ \frac{x^{2}}{2u} - \frac{x}{uy_0} \right] },
\end{eqnarray}
which will then determine $\mathcal{I}(x)$.
One is now left with the equation for $\mathcal{I}(x)$ reading
\begin{eqnarray}
 \displaystyle
 y_0 - u y_0 A(x) \frac{d \mathcal{I}(x)}{d x} = 0.
\end{eqnarray}
Plugging in the expression for $A(x)$ previously obtained and integrating on both sides one obtains:
\begin{eqnarray}
 \displaystyle
 \mathcal{I}(x) &=& \frac{1}{u}  \int^x dt \:  e^{\left[ \frac{-t^{2}}{2u} + \frac{t}{uy_0} \right]} 
\end{eqnarray}
The integral on the right-hand side is:
\begin{eqnarray}
 \displaystyle
 \int^x \! dt \: e^{\left[ \frac{-t^{2}}{2u} + \frac{t}{uy_0}\right] }& = &\sqrt{2u}\, e^{\frac{1}{2uy_0^2}}
 \int^{\frac{x}{\sqrt{2u}}-\frac{1}{\sqrt{2uy_0^2}}} d\tilde{t}  e^{-\tilde{t}^2} \nonumber\\
 &=&\frac{\sqrt{2u\pi}}{2} \, e^{\frac{1}{2uy_0^2}} \nonumber \\
&\times& erf \! \left[ \left( x-\frac{1}{y_0} \right) \frac{1}{\sqrt{2u}} \right]
 \end{eqnarray}
where the change of variables $\tilde{t}=\left( \frac{t}{\sqrt{2u}}-\frac{1}{\sqrt{2uy_0^2}}\right)$ has been made, and the lower limit of the last integral has been chosen to be zero, which requires to set a constant
$\bar{C}(u,y_0)$.  Hence
\begin{eqnarray}
 \displaystyle
 \mathcal{I}(x) &=&\sqrt{\frac{\pi}{2u}} e^{\frac{1}{2uy_0^2}} erf\left[ \left( x-\frac{1}{y_0} \right) \frac{1}{\sqrt{2u}} \right] 
\nonumber \\
 &+& \bar{C}(u,y_0).
 \end{eqnarray}
The exact solution $\displaystyle y_u(x)=A(x)\cdot \mathcal{I}(x)$ is given in Eq.\  (\ref{eq:gensolode_2}), where $\displaystyle C(u,y_0) = - \sqrt{\frac{2u}{\pi}}\bar{C}(u,y_0) e^{\frac{-1}{2uy_0^2}}$. 
%
%
%
%

\section{N-points continued fraction approximation \label{sec:appcf}}
We detail here how we have obtained the result of Eq.~(\ref{Eqn:cfa11}), or the order $\displaystyle \mathcal{O}(d^{3}_x)$ of the N-points limited order DE. \\
The starting point is Eq.~(\ref{Eqn:cf1}), which is differentiated with respect to the external potential, yielding:
\bea
\displaystyle
\label{Eqn:cfa1}
 \frac{\dk^{2} G(1,2;[\bar{\ep}])}{\dk \ep(6) \dk \ep(7) } &= & G_H^0(1,6) \frac{\dk G(6,2;[\bar{\ep}])}{\dk \bar{\ep}(7)} \nonumber \\
 &+&  G_H^0(1,7)  \frac{\dk G(7,2;[\bar{\ep}])}{\dk \bar{\ep}(6)} \nonumber \\
 & + & \int d3 G_H^0(1,3) \bar{\ep}(3) \frac{\dk^{2} G(3,2;[\bar{\ep}])}{\dk \bar{\ep}(6) \dk \bar{\ep}(7) } \nonumber \\
&+ & i \int d3 d5 W(3^{+},5)  G_H^0(1,3) \nonumber \\
&\times &\frac{\dk^{3} G(3,2;[\bar{\ep}])}{\dk \bar{\ep}(7) \dk \bar{\ep}(6) \dk \bar{\ep}(5)}.
\eea
Neglecting the term $\displaystyle \frac{\dk^{3} G(3,2;[\bar{\ep}])}{\dk \bar{\ep}(7) \dk \bar{\ep}(6) \dk \bar{\ep}(5) }$ and taking the limit $\varphi = 0$ yields:
\bea
\label{Eqn:cfa2}
 \frac{\dk^{2} G(1,2)}{\dk \ep(7) \dk \ep(6) } &=& G_H^0(1,6) \frac{\dk G(6,2;[\bar{\ep}])}{\dk \bar{\ep}(7)} \Big{|}_{\varphi=0} \nonumber \\
  &+& G_H^0(1,7) \frac{\dk G(7,2;[\bar{\ep}])}{\dk \ep(6)}\Big{|}_{\varphi=0} .
\eea
By substituting back Eq.~(\ref{Eqn:cfa2}) into Eq.~(\ref{Eqn:cf1}) we get:
\bea
\label{Eqn:cfa3}
\frac{\dk G(1,2;[\bar{\ep}])}{\dk \bar{\ep}(6)}\Big{|}_{\varphi=0} &=& G_H^0(1,6)G(6,2) \nonumber \\
&+& i \int d3d5 G_H^0(1,3) W(3^{+},5)  \nonumber \\ 
& \times & \ob G_H^0(3,6) \frac{\dk G(6,2;[\bar{\ep}])}{\dk \bar{\ep}(5)}\Big{|}_{\varphi=0} \nonumber \\
& +& G_H^0(3,5)\frac{\dk G(5,2;[\bar{\ep}])}{\dk \bar{\ep}(6)} \Big{|}_{\varphi=0} \cb.
\eea
The above equation can be recast in a compact way
\bea
\label{Eqn: cfa4}
 B_{xy} &=& B^{0}_{xy} + \sum_{qp} \gamma_{(xy)(qp)}B_{qp},
\eea
namely
%
%
%
\bea
\displaystyle
&& \frac{\dk G(1,2;[\bar{\ep}])}{\dk \bar{\ep}(6)} \Big{|}_{\varphi=0} = G_H^0(1,6)G(6,2) \nonumber \\
&+&  i \int d3d5d7  G_H^0(1,3) W(3^{+},5) G_H^0(3,6) \nonumber \\
 &\times&  \delta(7,6) \frac{\dk G(7,2;[\bar{\ep}])}{\dk \bar{\ep}(5)}\Big{|}_{\varphi=0} + \int d3 d5 d7G_H^0(1,3)   \nonumber \\ 
 &\times&  W(3^{+},5) G_H^0(3,5) \delta(7,6) \frac{\dk G(5,2;[\bar{\ep}])}{\dk \bar{\ep}(7)} \Big{|}_{\varphi=0}.
\eea
In the second term on the r.h.s. one can exchange, under the integral symbol, the indices 5 and 7, to obtain:
\bea
 \displaystyle
\label{Eqn:cfa5}
\frac{\dk G(1,2;[\bar{\ep}])}{\dk \bar{\ep}(6)} \Big{|}_{\varphi=0} &=& G_H^0(1,6)G(6,2) \nonumber \\
&+& i \int d3 d5 d7  G_H^0(1,3) W(3^{+},5) G_H^0(3,6) \nonumber \\
& \times &\delta(7,6) \frac{\dk G(5,2;[\bar{\ep}])}{\dk \bar{\ep}(7)}\Big{|}_{\varphi=0} \nonumber \\ 
& + & \int d3 d5 d7 G_H^0(1,3) W(3^{+},5) G_H^0(3,5) \nonumber \\
& \times & \delta(7,6) \frac{\dk G(5,2;[\bar{\ep}])}{\dk \bar{\ep}(7)} \Big{|}_{\varphi=0}. 
\eea
Let's now define the following quantities:
\bea
 \displaystyle 
&& \frac{\dk G(1,2;[\bar{\ep}])}{\dk \bar{\ep}(6)} \Big{|}_{\varphi=0} := g(1,6), \nonumber \\
&& \frac{\dk G(5,2;[\bar{\ep}])}{\dk \bar{\ep}(7)} \Big{|}_{\varphi=0} := g(5,7), \nonumber \\
&& G_H^0(1,6) G(6,2) := g_0(1,6), \nonumber \\
\label{Eqn:cfa5bis}
&& m(1,6;5,7) := i \int d3 G_H^0(1,3)W(3^{+},5)  \delta(7,6) \nonumber \\
&& \left[ G_H^0(3,6) + G_H^0(3,5) \right]. 
\eea
Recasting Eq.~(\ref{Eqn:cfa5}) with the new variables yields:
\be
\displaystyle
\label{Eqn:cfa6}
g(1,6) = g^{0}(1,6) + \int d5 d7 \, m(1,6;5,7) g(5,7)
\ee
we solve for $g$:
\bea
\displaystyle
\label{Eqn:cfa7}
\int d5 d7 \ob m(1,6;5,7) &-& \delta(1,5)\delta(7,6) \cb g(5,7) \nonumber \\
&+& g_0(1,6) = 0.
\eea
Defining
\bea
\displaystyle
 \label{Eqn:cfa8}
\ob  m(1,6;5,7) - \delta(1,5)\delta(7,6)  \cb =  \bar{m}(1,6;5,7),
 \eea
inserting this expression into Eq.~(\ref{Eqn:cfa7}):
 \be
 \displaystyle
 \label{Eqn:cfa9}
  \int d5 d7 \bar{m}(1,6;5,7) g(5,7) + g_0(1,6) = 0,
 \ee
and introducing the inverse of $\bar{m}$ one gets:
\bea
\displaystyle
&& \int d1 d6 d5 d7 \bar{m}^{-1}(8,9;1,6) \bar{m}(1,6;5,7) g(5,7) = \nonumber \\
&-& \int d1 d6 \bar{m}^{-1}(8,9;1,6) g_0(1,6) \\
&& g(8,9)  = - \int d1 d6  \bar{m}^{-1}(8,9;1,6) g_0(1,6).
\eea
Transforming back to the \textit{original} variables gives:
\be
\displaystyle
\label{Eqn:cfa10}
\frac{\delta G(1,2;[\bar{\ep}])}{\delta \bar{\ep}(6)}\Big{|}_{\varphi=0} = - \int d9d8 \bar{m}^{-1}(1,6;9,8) G_H^0(9,8)G(8,2)
\ee
and finally the Green's function reads:
\bea
\displaystyle
\label{Eqn:cfa11b}
 G(1,2)  & = & G_H^0(1,2) \nonumber \\
         &-& i \int d 5 d3 d8 d9 G_H^0(1,3) W(3^{+},5) \nonumber \\
         &\times& \bar{m}^{-1}(3,5;9,8) G_H^0(9,8)G(8,2)       
\eea
%
%
%
%
%
%
%

%
--------------------
----------------%

%
%
\end{document}